\documentclass[aps,prb,notitlepage,superscriptaddress,twocolumn]{revtex4-2}
\usepackage{natbib}
\usepackage{graphicx}
\usepackage{amssymb}
\usepackage{amsmath}
\usepackage{ifthen}
\usepackage{braket}
\usepackage{xcolor}
\usepackage{bm}
\usepackage{bbm}
\usepackage{comment}
\usepackage{siunitx}
\usepackage{float}
\usepackage{dcolumn}
\newcolumntype{d}[1]{D{.}{.}{#1}}
\usepackage{subfigure}

\newcommand{\dd}{\mathrm{d}}
\newcommand{\ii}{\mathrm{i}}
\newcommand{\ee}{\mathrm{e}}

\newcommand{\CM}{\mathrm{CM}}

\newcommand{\oneosc}{\mathrm{1osc}}
\newcommand{\fosc}{\mathrm{fosc}}

\newcommand{\brek}{\mathrm{br}}
\newcommand{\crit}{\mathrm{cr}}

\definecolor{garrosgreen}{rgb}{0.1, 0.4, 0.1}
\definecolor{dartmouthgreen}{rgb}{0.05, 0.5, 0.06}
\definecolor{duelferred}{rgb}{0.7, 0.2, 0.1}
\definecolor{cambridgeblue}{rgb}{0.1, 0.3, 1.0}
\definecolor{oxfordblue}{rgb}{0.05, 0.2, 0.7}

\begin{document}

\title{Retardation Effects in Atom-Wall Interactions }

\author{T. Das}
\affiliation{Department of Physics and LAMOR, Missouri University of Science and
Technology, Rolla, Missouri 65409, USA}

\author{C. A. Ullrich}
\affiliation{Department of Physics and Astronomy,
University of Missouri, Columbia, Missouri 65211, USA}

\author{U. D. Jentschura}
\affiliation{Department of Physics and LAMOR, Missouri University of Science and
Technology, Rolla, Missouri 65409, USA}

%
%
\begin{abstract}
The onset of retardation effects in atom-wall interactions is
studied. It is shown that the transition range from the
$1/z^3$ short-range (van der Waals) interaction to the
$1/z^4$ long-range (Casimir) retarded interaction critically
depends on the atomic properties and
on the dielectric function of the material.
For simple non-alkali atoms (e.g., ground-state hydrogen and
ground-state helium) interacting
with typical dielectric materials such as intrinsic silicon,
the transition to the retarded regime is shown
to proceed at a distance of about $10\,{\rm nm}$ (200 Bohr radii).
This is much shorter than typical characteristic
absorption wavelengths of solids.
Larger transition regimes are obtained
for atoms with a large static polarizability
such as metastable helium.
We present a simple estimate for the critical distance,
$z_{\crit} = 137 \, \sqrt{\alpha(0)/Z}$
atomic units, where $\alpha(0)$ is
the static polarizability (expressed
in atomic units) and $Z$ is the number of electrons
of the atom.
\end{abstract}

\maketitle

\tableofcontents

%
%
\section{Introduction}

Dispersion forces between spatially well-separated microscopic systems are
important for phenomena such as atom-surface scattering, physisorption, the
structure of soft matter and 2D layered materials, and many applications
\cite{FrEtAl2010,WoEtAl2016}. In this context, it is well known that atom-atom
interactions undergo a transition from a short-range van der Waals ($1/R^6$) to
a retarded long-range ($1/R^7$) behavior, where $R$ is the interatomic distance
(see Ref.~\cite{BeLiPi1982vol4} and Chaps.~4 and 11 of
Ref.~\cite{JeAd2022book}).  For atom-wall interactions, the asymptotic
behavior changes from $1/z^3$ for short-range to $1/z^4$ in the long-range
limit (see, e.g., Ref.~\cite{LaDKJe2010pra}), due to a process called
retardation. The interpolating formula has been
given in Eqs.~(18) and (21) of Ref.~\cite{AnPiSt2004}
(see also Ref.~\cite{Je2024multipole}).
However, the precise nature of this transition is
less well characterized in the literature.
From Fig.~3 of Ref.~\cite{AnPiSt2004},
it is evident that the interaction of ${}^{87}{\rm Rb}$
atoms with a sapphire surface starts to
substantially deviate
from the $1/z^3$ short-range asymptotics
in the range $z \sim 30 \, {\rm nm} \approx 600 \, a_0$,
where $a_0$ is the Bohr radius.
For the example of metastable helium
(in the triplet state)
interacting with a gold surface,
estimates for the transition region to the
retarded regime have
been indicated in the range of
$z \leq 150 \, {\rm nm} \approx 3000 \, a_0$
in the text following Eq.~(3)
in Sec.~III of Ref.~\cite{CaKlMoZa2005},
and in Sec.~16.3.4 and Sec.~16.4.2 of
Ref.~\cite{BoKlMoMo2009}.
Here, we concentrate on metastable 
helium in the triplet $2 {}^3S_1$ state,
which has a radiative lifetime of about $7800$\,s
and is thus sufficiently long-lived to probe
atom-surface interactions in detail,
including quantum reflection studies~\cite{DrDK2003}. By contrast, while the 
$2 {}^1S_0$ state also is metastable 
(dipole decay to the ground state is not allowed),
its lifetime is much shorter,
of the order of only $2 \times 10^{-2}$\,s.
In this paper, we aim to provide clarity and give
both simple estimates and precise numerical results
that show the onset and
spatial range of the transition regime
between van der Waals and Casimir-Polder
interactions. The dependence of the
transition region on the atomic species
and on the dielectric function of the surface
is also studied.

Intuitively, we can understand the onset of
retardation as follows: Atom-wall
interactions happen due to the exchange
of virtual photons. If an exchange photon picks up
a nonnegligible phase (of order unity) on its way from
the atom to the wall and back, retardation needs
to be taken into account (Chap. 5 of Ref.~\cite{JeAd2022book}).
The phase of a characteristic photon is given as
$\Delta \phi = k_{\rm ch} \, z$,
where $k_{\rm ch}$ is the wave vector corresponding to
a characteristic resonance excitation of the
atom (or solid).
The condition $\Delta \phi \sim 1$ leads to
$z \sim 1/k_{\rm ch} = \lambda_{\rm ch}/(2 \pi)$,
where $\lambda_{\rm ch}$ is the
characteristic wavelength.
For simple atomic systems such as (atomic)
hydrogen or helium (in their ground states),
the characteristic excitation wavelength
is $\lambda_{\rm ch} = \hbar c/E_h$,
where $E_h = \alpha^2 m c^2$ is the
Hartree energy (where $\alpha$ is the fine-structure
constant, $m$ is the electron mass,
$c$ is the speed of light, 
and the subscript $h$ stands for Hartree). 
Hence, {\em a priori},
we can expect retardation effects to become important
when the atom-wall distance is
of the order of a Hartree wavelength $\lambda_h$,
\begin{equation}
\label{hegel}
z \sim \lambda_h = \frac{\hbar c}{E_h} =
\frac{a_0}{\alpha} = 7.25 \, {\rm nm} = 137 \, {\rm a.u.}\,,
\end{equation}
where $a_0$ is the Bohr radius,
which is the unit of length in the
atomic unit system. Throughout this article, we use the 
acronym (a.u.) for quantities given in atomic units.
We note that $\lambda_h$ is, purely parametrically,
of the same order as optical wavelengths,
but typical optical wavelengths in the
visible spectrum are longer than $\lambda_h$;
the UV spectrum ends at about 400\,nm.
Hence, one might ask whether or not
large prefactors could shift the parametric
estimate~\eqref{hegel}.

Here, we demonstrate that an
extended distance scale for the
nonretarded interaction may be observed
for special atoms with an excessively large
static polarizability, but that
retardation sets in at much shorter
length scales commensurate with
Eq.~\eqref{hegel} (in typical cases,
about $10 \, {\rm nm} \approx 200 \, {\rm a.u.}$)
for many simple atomic systems.
For example, we demonstrate by explicit numerical
calculations that the atom-wall interaction
of ground-state helium atoms
undergoes a transition to the retarded regime
much earlier, at length scales commensurate with 
$z \sim \lambda_h$.
Variations of the onset of the
retarded regime with the atomic system
are also discussed.

This paper is organized as follows:
We discuss the interpolating formula
for the transition from the short-range
to the long-range regime in Sec.~\ref{sec2},
with a special emphasis on interactions
of hydrogen and helium with a silicon surface.
Other elements are discussed in Sec.~\ref{sec3}.
Atomic units are used throughout
unless indicated otherwise ($\hbar = e = 1$,
$\epsilon_0 = 1/(4 \pi)$, $c = 1/\alpha$,
where $\alpha$ is the fine-structure constant).
We provide mini-reviews of applicable distance
ranges in Appendix~\ref{appa},
parameters of our helium calculations in 
Appendix~\ref{appb}, and of the dielectric function of intrinsic silicon in
Appendix~\ref{appc}. A derivation of
the Thomas--Reiche--Kuhn (TRK) sum rule
for metastable reference states \cite{ReTh1925zahl,Ku1925}
is presented in Appendix~\ref{appd}.

%
%
\section{Hydrogen and Helium on Silicon and Gold}
\label{sec2}

We start from an interpolating expression for the
atom-wall interaction, which reduces to the
$1/z^3$ short-range interaction for small
atom-wall distance and to the
$1/z^4$ long-range interaction for large
distance. The relevant formula
is given in Eqs.~(18) and~(21)
of Ref.~\cite{AnPiSt2004},
\begin{align}
\label{Vatom}
\mathcal{E}(z) =& \;
-\frac{\alpha^3}{2 \pi} \int\limits_0^\infty \dd\omega\,\omega^3 \,
\alpha(\ii \omega)
\int\limits_1^\infty\dd p
\ee^{- 2 \, \alpha\, p  \, \omega \, z }
H( \epsilon( \ii \omega), p )\, ,
\end{align}
where
\begin{equation}
\label{defH}
H(\epsilon, p)= \frac{\sqrt{\epsilon-1+p^2}-p}
{\sqrt{\epsilon-1+p^2}+p}
+(1-2p^2)
\frac{\sqrt{\epsilon-1+p^2}-p \, \epsilon}
{\sqrt{\epsilon-1+p^2}+p \, \epsilon}.
\end{equation}
Here, $\alpha(\ii \omega)$ is the dynamic
(dipole) polarizability of the atom at imaginary
driving frequency,
and $\epsilon(\ii \omega)$ is the
dielectric function of the solid at imaginary angular
frequency.
For the material of the solid (intrinsic silicon), we assume the interpolating
model of the temperature-dependent dielectric function
recently discussed in Ref.~\cite{MoEtAl2022}
for intrinsic silicon (with slight modifications).
The parameters are reviewed in Appendix~\ref{appc}.
We also study gold, employing a simple plasma model
for its dielectric function for definiteness,
and a modified model discussed in
Eq.~(13.46) of Ref.~\cite{BoKlMoMo2009}.

In the current section, we focus
on atomic hydrogen and helium.
For hydrogen, we employ the following
formula for the dipole polarizability in the
non-recoil approximation (infinite nuclear mass), which is sufficient
for the accuracy required in the current investigation,
\begin{equation}
\label{alphaH}
\alpha_{\rm H}(\omega) =
Q_{\rm H}(\omega) +
Q_{\rm H}(-\omega) \,,
\end{equation}
where
\begin{equation}
Q_{\rm H}(\omega) = \frac13 \,
\left< 1S \left| \vec r \,
\frac{1}{H - E_{1S} + \hbar \omega}
\, \vec r \right| 1S \right> \,,
\end{equation}
where $E_{1S}$ is the ground-state energy
of hydrogen, $H$ is the Schr\"{o}dinger--Coulomb
Hamiltonian, and the scalar product is understood
for the two position operators.
According to Eq.~(4.154) of Ref.~\cite{JeAd2022book},
the dipole matrix element can be expressed as follows,
\begin{equation}
\label{Q1H}
Q^{({\rm H})}(\omega) = \frac{2 t^2 \, p(t)}{3 \, (1-t)^5 \, (1+t)^4}
+ \frac{256 \,t^9 \, f(t)}{3\,(1+t)^5 \,(1-t)^5} \,,
\end{equation}
where the photon energy is parameterized by the $t$ variable,
$t = t(\omega) = (1 + 2 \omega)^{-1/2}$.
The polynomial $p(t)$ incurred in Eq.~\eqref{Q1H}
is
\begin{equation}
p(t) = 3 - 3 t - 12 t^2 + 12 t^3 + 19 t^4 - 19 t^5
- 26 t^6 - 38 t^7 \,.
\end{equation}
The function $f(t)$ is a complete hypergeometric function,
\begin{equation}
f(t) = {}_2 F_1(1, -t, 1-t, \xi ) \,,
\end{equation}
and $\xi = (1-t)^2/(1+t)^2$.

For helium, we use an approach based
on a fully correlated basis set,
using exponential basis functions~\cite{Ko1999,Ko2000,Ko2002pra}.
We employ a L\"{o}wdin decomposition
of the overlap matrix (see Appendix~J of Ref.~\cite{Pi2013})
and use extended-precision arithmetic in the {\tt julia}
language~\cite{julialang,Ba2015,BeEdKaSh2017}
in order to avoid loss of numerical
precision in intermediate steps of the
calculation. The {\tt HTDQLS} algorithm~\cite{NoLuStJe2017cpc}
is used to diagonalize the overlap matrix
in arbitrary precision. 
Trial functions for $S$ states are of the explicitly correlated 
form $\exp(-a r_1 - b r_2 - c r_{12})$
with an appropriate symmetrization 
for particle interchange (positive sign for singlet 
states, negative sign for triplet states).
Here, $r_1$ and $r_2$ are the distances of 
the two electrons from the helium nucleus,
and $r_{12}$ is the interelectron distance.
For $P$ states, 
trial functions are of the explicitly correlated
form $x_1^i \, \exp(-a r_1 - b r_2 - c r_{12})$,
where $\vec r_1 = \sum_{i=1}^3 x_i \hat e_i$
in the Cartesian basis, again 
with an appropriate symmetrization
for particle interchange.
For the calculation of the matrix elements
of the Hamiltonian, and the dipole transition
matrix elements, one needs the master integrals
given in Chap.~13 of Ref.~\cite{JeAd2022book}.
Further details are relegated to Appendix~\ref{appb}.

\begin{figure}[t]
\begin{center}
\includegraphics[width=0.93\linewidth]{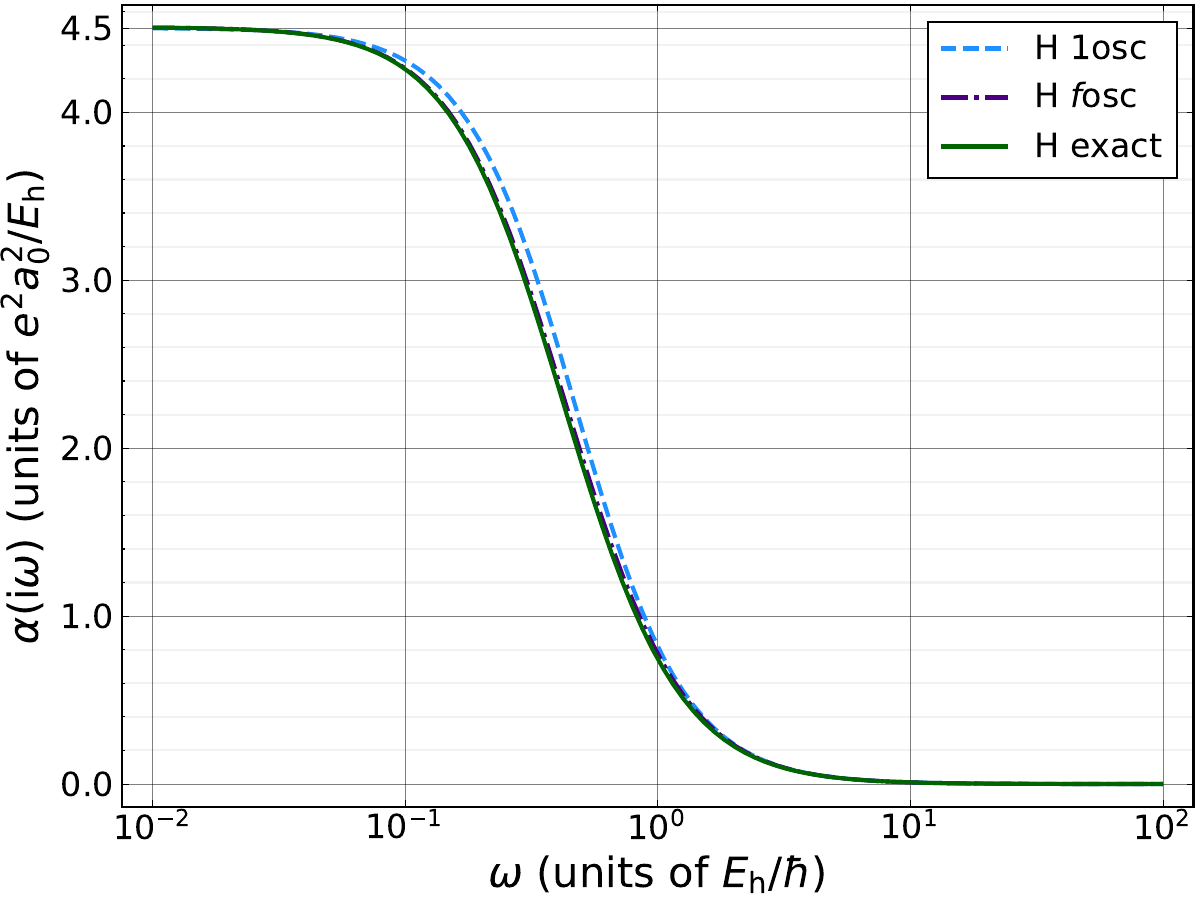}
\end{center}
\caption{\label{fig1} Dynamic (dipole) polarizability of
atomic hydrogen, $\alpha(\ii \omega)$, as a function
of the imaginary driving frequency. The exact
values obtained from Eq.~\eqref{alphaH} are compared with the
few-oscillator-strength based model (fosc) described in Appendix B of
Ref.~\cite{MoEtAl2022}. The first 30 oscillator strengths
and their corresponding transition energies are collected from Table 4 of
Ref.~\cite{WiFu2009} for the evaluation,
which yields a maximum relative error of $4.16\%$.
For comparison, the single-oscillator model, given in
Eq.~\eqref{alpha1osc}, is also plotted.
The atomic polarizability is given in atomic units,
i.e., in units of $e^2 a_0^2/E_h$, where $e$ is the electron charge,
$a_0$ is the Bohr radius, and $E_h$ is the Hartree energy.}
\end{figure}

\begin{figure}[t]
\begin{center}
\includegraphics[width=0.93\linewidth]{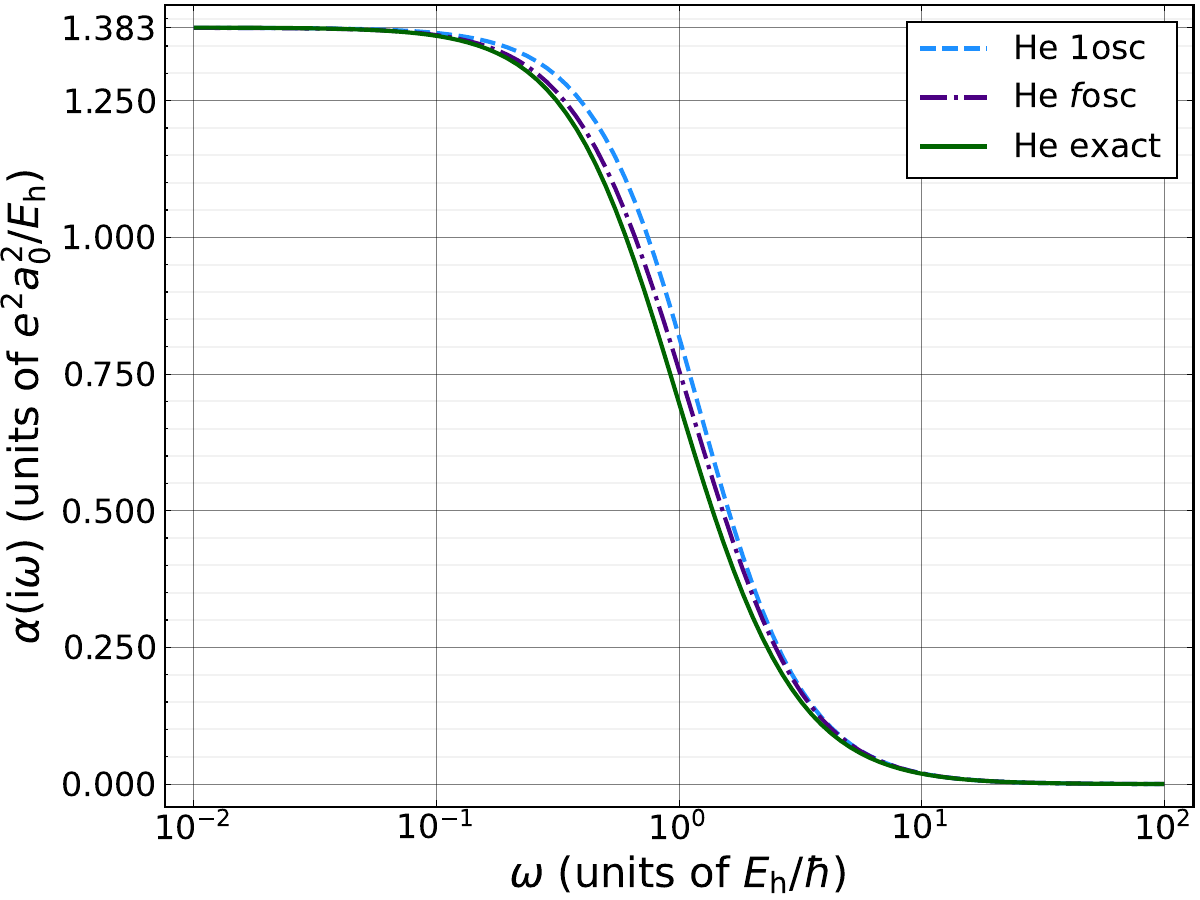}
\end{center}
\caption{\label{fig2} Dynamic (dipole) polarizability of
(ground-state) atomic helium as a function of imaginary driving frequency. The values
obtained from the exact approach based on a fully correlated basis set are
compared with the oscillator-strength based model. Oscillator strengths for
excited states from $n=2$ to $n=10$, and their corresponding transition
energies are collected from Table 14 of Ref.~\cite{WiFu2009} for the
evaluation, which yields a maximum relative error of $12.14\%$.
For comparison, the single-oscillator model 
[Eq.~\eqref{alpha1osc}] is also plotted.}
\end{figure}

\begin{figure}[t]
\begin{center}
\includegraphics[width=0.93\linewidth]{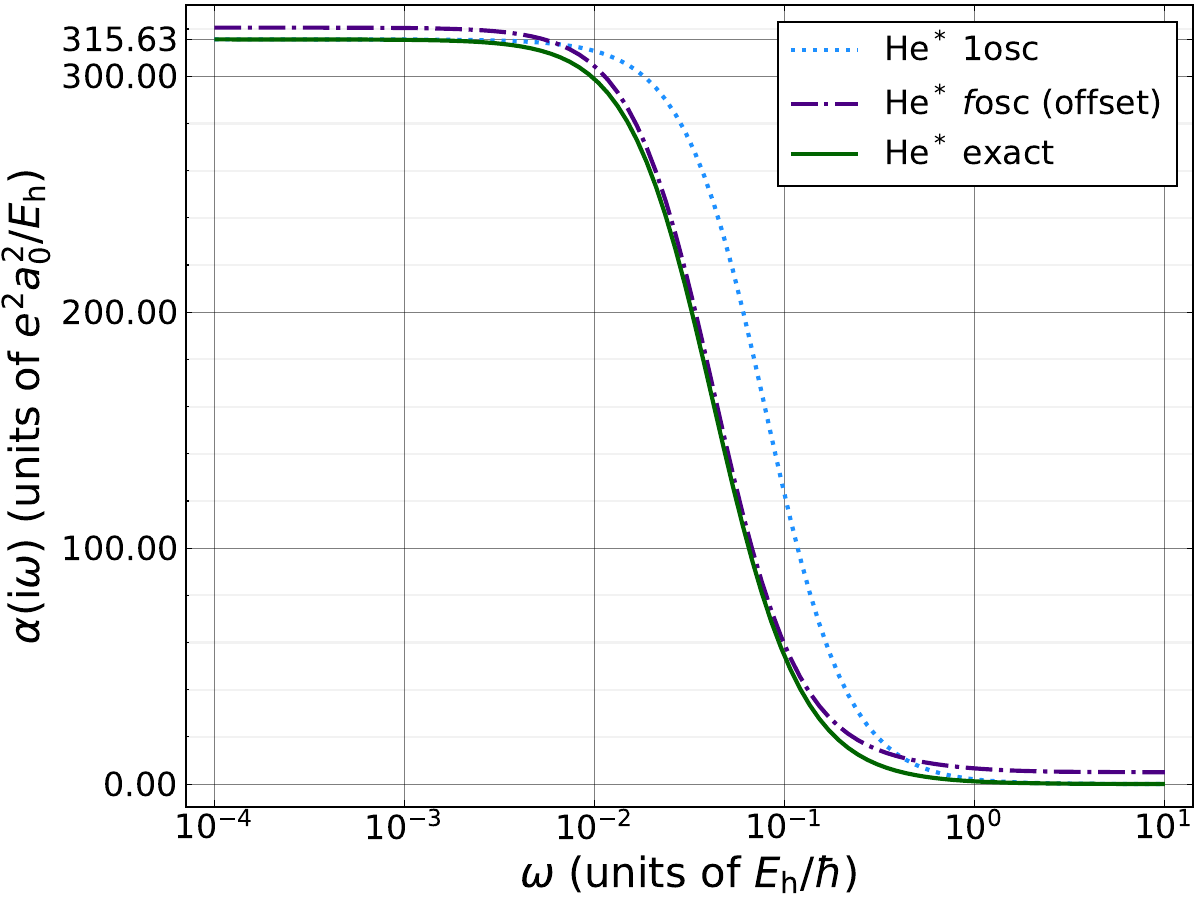}
\end{center}
\caption{\label{fig3} Same as Figs.~\ref{fig1} and~\ref{fig2},
but for metastable helium (He$^*$) in the $2 {}^3S_1$ reference
state. Because the two curves for the FOSC model and the exact polarizability
almost overlap, the curve for the FOSC model is shifted upwards by a
constant offset of $+5.0\,{\rm a.u.}$ in order to make the curves visually
discernible.}
\end{figure}

\begin{figure}[t]
\begin{center}
\includegraphics[width=0.93\linewidth]{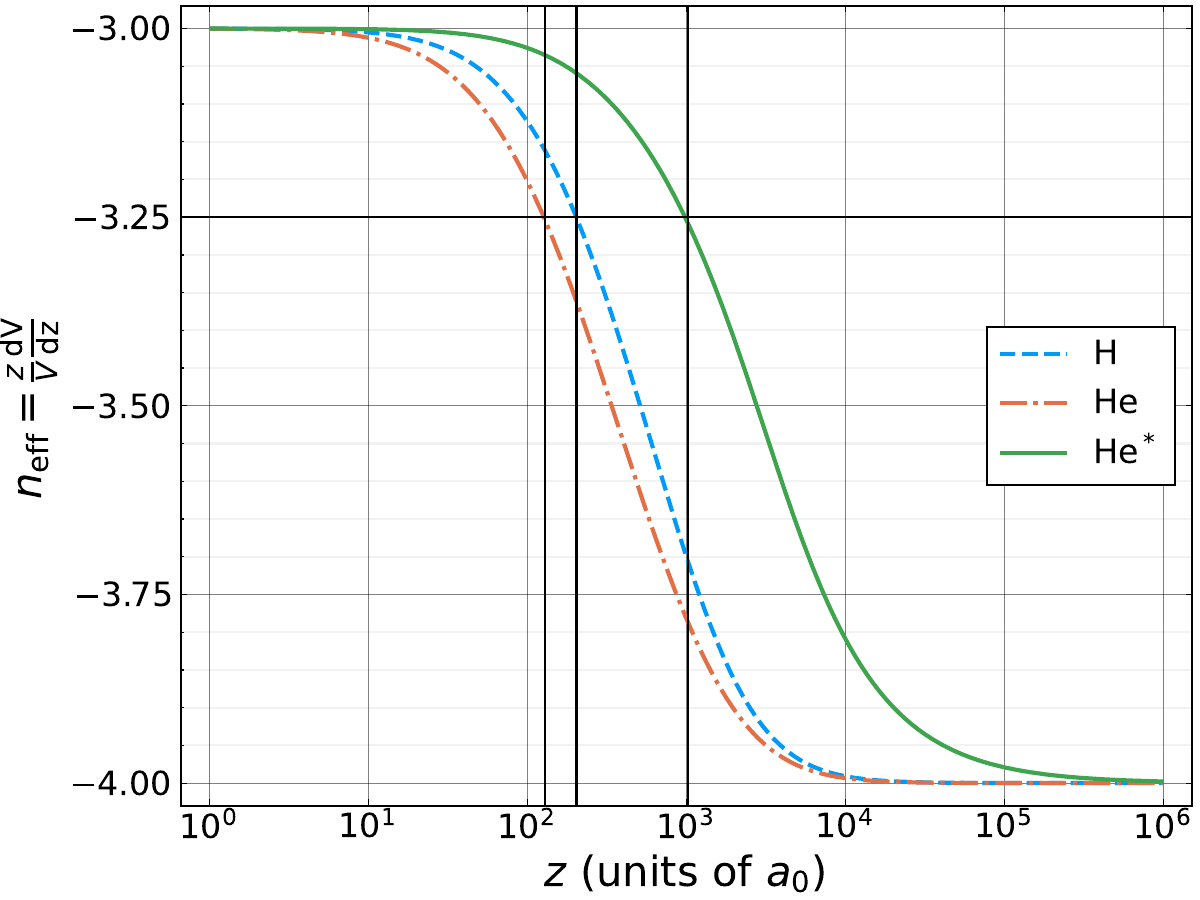}
\end{center}
\caption{\label{fig4}
Change in the effective exponent $n_{\rm eff}$ for atom-wall interactions due to
the transition from the short-range (van der Waals) to the long-range (Casimir) regime
for hydrogen, ground-state helium and metastable 2$^3S_1$
helium, interacting with intrinsic
silicon. To this end, the atom-wall potential is numerically
evaluated and the effective exponent
is calculated via Eq.~\eqref{neff} as a function of the atom-wall separation.
The Clausius-Mossotti (CM)
model described in Ref.~\cite{MoEtAl2022} is used for the dielectric
function of intrinsic silicon, with slightly
modified parameters (see Table~\ref{table1}).
The exact dynamic polarizability is used for all atoms.}
\end{figure}

For the evaluation of the dynamic polarizability,
two other approaches have been discussed in the
literature, namely, the single-oscillator
model~\cite{BaKlMoZa2004,CaKlMoZa2005}
(henceforth referred to by the acronym 1osc)
and the few-oscillator model~\cite{MoEtAl2022}
(henceforth referred to by the acronym fosc).
The single-oscillator model, asymptotically matched
to the static ($\omega \to 0$)
and ultraviolet ($\omega \to \infty$) limits,
reads as follows (in atomic units),
\begin{equation}
\label{alpha1osc}
\alpha_{\oneosc}(\ii \omega)
= \frac{Z}{ \omega^2 + Z/\alpha(0)}
= \frac{Z}{ \omega^2 + \omega_{\crit}^2} \,.
\end{equation}
Here, $\alpha(0)$ is the static polarizability,
$Z$ is the number of electrons,
and the critical frequency
is $\omega_{\crit} = \sqrt{ Z/\alpha(0)}$
(this frequency will be important for our considerations
in Sec.~\ref{sec3}).
The formula~\eqref{alpha1osc} has the correct
static limit [$\alpha( \ii \omega = 0) = \alpha(0)$].
The correct ultraviolet limit is
also obtained in view of the asymptotic relation
$\alpha(\ii \omega) \to Z/\omega^2$,
which fulfills
the Thomas--Reiche--Kuhn (TRK) sum rule
\cite{ReTh1925zahl,Ku1925}. This sum rule remains
valid for metastable reference states
(see Appendix~\ref{appd}).
Values of $\alpha(0)$ have been tabulated in
Ref.~\cite{ScNa2019} for all elements
with nuclear charge numbers $1 \leq Z \leq 120$.
(Remark: It is also possible to match the
single-oscillator model against the
van der Waals coefficient of the atomic
dimer system~\cite{BrEtAl2002,Ta1969},
but in this case, one fails to fulfill the
TRK sum rule in the ultraviolet region.
Here, we use the functional form
given in Eq.~\eqref{alpha1osc}.)

As an intermediate between the exact calculation
of the dynamic polarizability and the
single-oscillator model, the few-oscillator
model has recently been discussed in
Appendix~B of Ref.~\cite{MoEtAl2022}.
Let us assume that a finite number of oscillator
strengths $f_n$ are known ($n \in \{1, \ldots, N\}$),
with corresponding resonance frequencies $\omega_n$.
The few-oscillator strength model reads as follows,
\begin{subequations}
\label{alphafosc}
\begin{align}
\label{alphafosc_a}
\alpha_{\fosc}(\ii \omega) =& \; \sum_{n=1}^N
\frac{f_n}{\omega^2 + \omega_n^2}
+ \frac{1}{\omega^2 + \omega_c^2} \,
\left( Z - \sum_{n=1}^N f_n \right) \,,
\\
\label{alphafosc_b}
\left(\omega_c\right)^2 = & \;
\frac{ Z - \sum_{n=1}^N f_n }{ \alpha(0) - \sum_{m=1}^N f_m/\omega_m^2 } \,.
\end{align}
\end{subequations}
Here, $\omega_c$ describes the typical scale of
virtual excitations into the continuum.
One collects a number $N$ of oscillator strengths
[first term on the right-hand side of Eq.~\eqref{alphafosc_a}]
and approximates the completion of the spectrum
by including the
second term on the right-hand side of Eq.~\eqref{alphafosc_a}.
The choice of the frequency $\omega_c$ in Eq.~\eqref{alphafosc_b}
ensures that the correct static limit $\alpha(0)$ is reproduced.
From tables (e.g., Ref.~\cite{WiFu2009}),
it is possible to collect at least $N=9$ oscillator strengths
to the lowest excited states for typical atomic
species.

In Figs.~\ref{fig1} and~\ref{fig2},
and Fig.~\ref{fig3} for metastable helium,
we show that the oscillator-strength-based
approach used in Ref.~\cite{MoEtAl2022},
which is enhanced by matching the static
polarizability and the ultraviolet limits
with accurate limits with exact results,
yields numerical data for the dynamic polarizability
of hydrogen and helium which are in good agreement
with the exact results.
A comparison of the single-oscillator
model to the few-oscillator--strength
model illustrates the gradual improvement
achieved by including more known oscillator
strengths. Another observation is as follows:
The presence of resonances
due to transitions to other bound--state energy
levels has a tendency to lower the
curve of $\alpha(\ii \omega)$ upon the inclusion
of more bound-state resonances
as compared to the single-oscillator model,
i.e., one has $\alpha_{\oneosc}(\ii \omega) >
\alpha_{\fosc}(\ii \omega) > \alpha(\ii \omega)$.
Hence, the single-oscillator model tends
to overestimate the polarizability at finite
excitation frequencies, while reproducing
the correct limit for very high frequencies.

In order to gauge the transitions from
the short-range to the long-range regime,
we use the effective ``local'' power coefficient
\begin{equation}
\label{neff}
n_{\rm eff} = \frac{z}{V(z)} \frac{\dd V(z)}{\dd z} =
\frac{\dd \ln(| V(z) |)}{\dd \ln(z)}  \,.
\end{equation}
It evaluates to exactly $n$ when $V(z) = V_0 z^n$.
By the logarithm of the potential,
we understand the
logarithm of the numerical value (reduced quantity)
of the potential, expressed in atomic units.

\begin{figure}[t!]
\begin{center}
\includegraphics[width=0.93\linewidth]{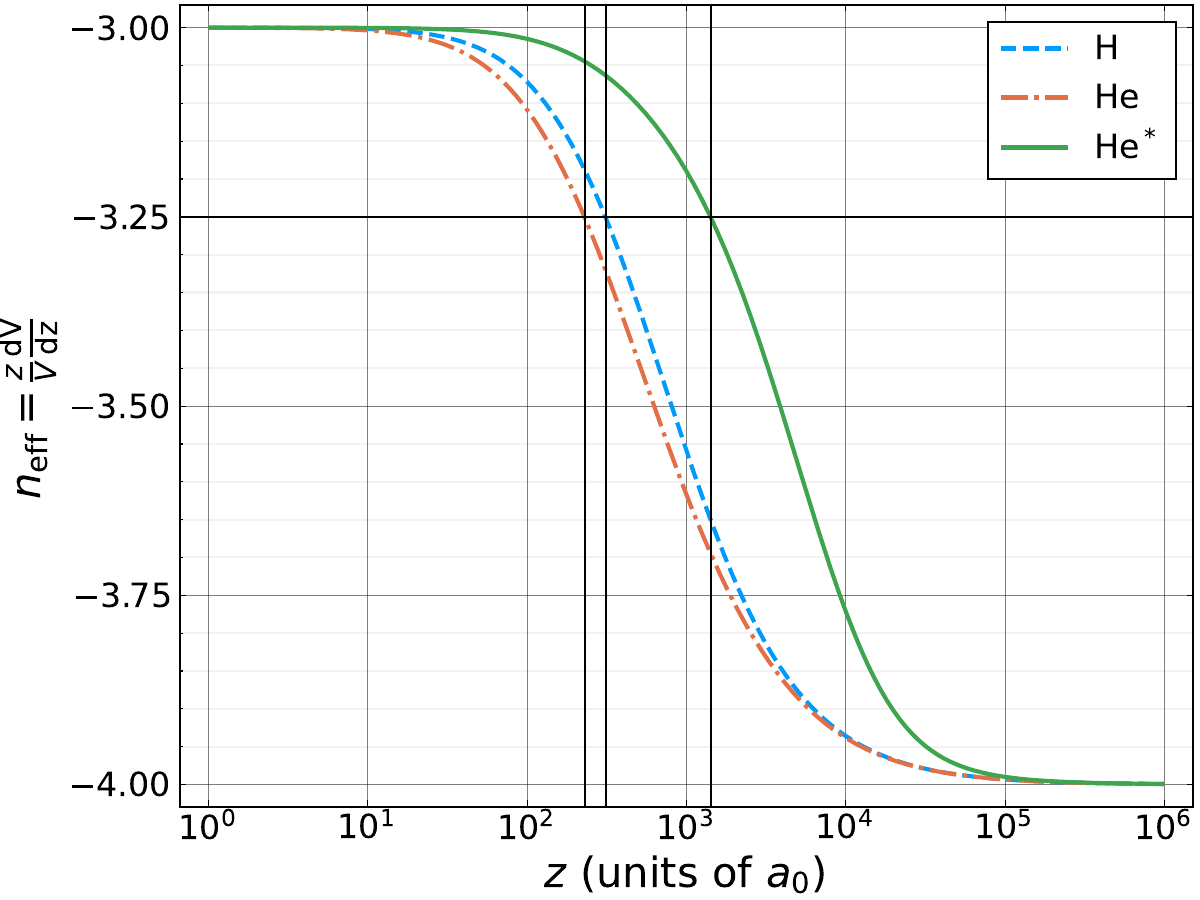}
\end{center}
\caption{\label{fig5}
Breakdown of the short-range asymptotics
for the atom-wall interaction for hydrogen interacting with gold,
as described by the plasma model~\eqref{plasma_model}.
Otherwise, the figure is analogous to
Fig.~\ref{fig4}.}
\end{figure}

The dependence of the effective exponent $n_{\rm eff}$ on the
atom-wall distance $z$ is shown in
Figs.~\ref{fig4} and~\ref{fig5} for the interaction of
H, He and H$^*$ with silicon and gold, respectively.
Let us define the break-down point $z_{\brek}$ for the
short-range expansion to be the distance
where the effective exponent $n_{\rm eff}$
reaches the value $n_{\rm eff} = -3.25$, which is 25\% of the
way between the asymptotic short-range value ($n_{\rm eff} = -3$)
and the long-range value ($n_{\rm eff} = -4$).
This definition, while arbitrary to some extent, captures the essence of the transition between the two regimes.
In addition to the substantial deviation
of the effective exponent $n_{\rm eff}$
from the value $n_{\rm eff} = -3$ at the break-down point,
we have checked that the relative deviation
of the atom-surface potential $V(z)$ from the
short-range estimate~\eqref{nonretard},
parametrized by the function
$D(z) = [V(z) - (-C_3/z^3)]/V(z)$,
is at least 35\,\% at $z = z_{\brek}$, further
validating the sensibility of our definition.

From Fig.~\ref{fig4}, one reads off
the following values for interactions with
intrinsic silicon,
\begin{subequations}
\label{zbreakSi}
\begin{align}
z_{\brek}({\rm H}; {\rm Si}) \approx & \; 203 \, {\rm a.u.} \,,
\\
z_{\brek}({\rm He}; {\rm Si}) \approx & \; 126 \, {\rm a.u.} \,,
\\
z_{\brek}({\rm He}^*; {\rm Si}) \approx & \;  1033 \, {\rm a.u.} \,.
\end{align}
\end{subequations}
When using the single-oscillator model, the
values change into
\begin{subequations}
\label{zbreakSi_single_osc}
\begin{align}
z_{\brek}({\rm H}; \oneosc; {\rm Si}) \approx & \; 194 \, {\rm a.u.} \,,
\\
z_{\brek}({\rm He}; \oneosc; {\rm Si}) \approx & \; 117 \, {\rm a.u.} \,,
\\
z_{\brek}({\rm He}^*; \oneosc; {\rm Si}) \approx & \; 674 \, {\rm a.u.} \,.
\end{align}
\end{subequations}
For the fosc model, the values of
$z_{\brek}$ are in between the values
for the exact polarizabilities and those
for the 1osc model, namely,
$200 \, {\rm a.u.}$, $121 \, {\rm a.u.}$, and $1005 \,{\rm a.u.}$,
respectively, for H, He, and He$^*$.

Another example is the calculation of
$z_{\brek}$ for interactions with gold,
where we use the plasma model,
\begin{equation}
\label{plasma_model}
\epsilon(\ii \omega) = 1 + \frac{\omega_{\rm pl}^2}{\omega^2} \,,
\end{equation}
where $\omega_{\rm pl}$ is the plasma frequency.
For the plasma frequency $\omega_{\rm pl}$,
we use the same value as advocated in Ref.~\cite{BoKlMoMo2009},
namely, $9 \, {\rm eV}$.
The dielectric function of the plasma model
diverges in the limit $\omega \to 0$,
which implies that the long-range limit
of the interaction with a gold surface
is the same as for a perfect conductor (see also
Ref.~\cite{Je2024multipole}).
The dielectric function of gold, approximated by the
plasma model, is strongly peaked for very low
frequency; it constitutes a cursory approximation.
Because of the strong emphasis on very low
virtual photon frequencies, we can expect
$z_{\brek}$ to be exceptionally large
as compared to other materials (see also
the discussion in Sec.~\ref{sec3}).

Figure \ref{fig5} shows the breakdown of the short-range
expansion for hydrogen, ground-state and metastable
helium, for interactions with gold.
One reads off the values
\begin{subequations}
\label{zbreakAu}
\begin{align}
z_{\brek}({\rm H}; {\rm Au}) \approx & \; 309 \, {\rm a.u.} \,,
\\
z_{\brek}({\rm He}; {\rm Au}) \approx & \; 228 \, {\rm a.u.} \,,
\\
z_{\brek}({\rm He}^*; {\rm Au}) \approx & \; 1414 \, {\rm a.u.} \,.
\end{align}
\end{subequations}
For the single-oscillator model, one obtains
\begin{subequations}
\label{zbreakAu_single_osc}
\begin{align}
z_{\brek}({\rm H};    \oneosc; {\rm Au}) \approx & \; 297 \, {\rm a.u.} \,,
\\
z_{\brek}({\rm He};   \oneosc; {\rm Au}) \approx & \; 217 \, {\rm a.u.} \,,
\\
z_{\brek}({\rm He}^*; \oneosc; {\rm Au}) \approx & \; 892 \, {\rm a.u.} \,.
\end{align}
\end{subequations}
For the fosc model, the values of
$z_{\brek}$ are in between the values
for the exact polarizabilities and those
for the 1osc model, namely,
$306 \,{\rm a.u.}$, $223 \,{\rm a.u.}$, and $1405 \,{\rm a.u.}$
respectively, for H, He, and He$^*$.

The break-down distance depends quite substantially
on the atomic system. These observations raise the question
of the general dependence of the breakdown
of the short-range expansion on the atomic
species, and on the properties of the solid.

We had already mentioned that the plasma model of
gold leads to exceptionally large values of
$z_{\brek}$. This can be ramified: For example,
the modified plasma model given
in Eq.~(13.46) of Ref.~\cite{BoKlMoMo2009}
adds additional terms 
which modify the plasma model 
given in Eq.~\eqref{plasma_model} 
for higher frequencies,
while the leading asymptotics 
for $\omega \to 0$ are unmodified.
The addition of further terms 
shifts the dominant contribution to the 
integrals to higher frequencies $\omega$,
mimicking larger values of $\omega_{\rm cr}$
and leading to smaller
values of $z_{\brek}$. (One notices that 
$\omega_\crit$ is proportional to $\sqrt{Z/\alpha(0)}$,
while $z_\brek$ can be estimated to be 
proportional to $\sqrt{\alpha(0)/Z}$,
according to Eq.~\eqref{zcritres}.)
This is indeed confirmed.
When using the exact polarizabilities and
the modified plasma model given
in Eq.~(13.46) of Ref.~\cite{BoKlMoMo2009}, the
results given in Eq.~\eqref{zbreakAu} change into
$201 \, {\rm a.u.}$, $123 \, {\rm a.u.}$, and
$1311 \, {\rm a.u.}$, respectively,
for H, He and He$^*$.
The relative change as compared to the 
simple plasma model given in Eq.~\eqref{plasma_model} is
smallest for metastable helium;
this is due to emphasis on smaller excitation
frequencies (small value of $\omega_{\rm cr}$
for metastable helium, see Fig.~\ref{fig3}), 
where the simple plasma model given in
Eq.~\eqref{plasma_model} and the modified 
plasma model given in Eq.~(13.46) of Ref.~\cite{BoKlMoMo2009}
share the same asymptotics.

%
%
\section{Other Elements}
\label{sec3}

The question of the dependence
of the break-down distance $z_{\brek}$
on the atomic species is made more urgent by
the observation that more complex atoms
with occupied inner shells typically have
a much larger static polarizability~\cite{ScNa2019},
and much smaller typical excitation energies
(at least to the first excited states,
see Ref.~\cite{NISTASD} for a compilation).
One might think that the
smaller (lowest) excitation energies
of more complex atoms could imply a much narrower
functional form of the dynamic polarizability
$\alpha(\ii \omega)$ for more complex
atoms, and hence, a drastic extension
of the nonretarded $1/z^3$ short-range regime.
However, one could also counter-argue that more
complex atoms also possess transitions to
much higher excited states. Hence, one could
argue that these higher-energy virtual transitions
might lower the distance range for the onset
of retardation.

The discussion of other atomic species
is made easier by investigating the
general structure of the atom-surface interaction
integral given in Eq.~\eqref{Vatom}.
In order to estimate how far the nonretarded
approximation is valid, let us start from the
regime of not excessively large $z$.
In this case, the exponential suppression
factor $\exp(-\alpha \omega p z)$ is
not very pronounced, and the dominant
integration region comes from large $p$.
We expand $H( \epsilon( \ii \omega), p )$ for large $p$
with the result,
\begin{equation}
H( \epsilon( \ii \omega), p ) \approx
2 p^2 \, \frac{\epsilon(\ii \omega) -1}%
{\epsilon(\ii \omega) + 1} \,,
\end{equation}
commensurate with the leading term recorded in
Eq.~(22) of Ref.~\cite{JeMo2023blog}.
One then carries out the integral over $p$ in
Eq.~\eqref{Vatom} and obtains the approximate
formula
\begin{align}
\label{expo_suppress}
\mathcal{E}(z) \approx & \;
-\frac{1}{4 \pi z^3} \int\limits_0^\infty
\ee^{- 2 \, \alpha\,  \omega \, z }
\alpha(\ii \omega) \,
\frac{\epsilon(\ii \omega) -1}%
{\epsilon(\ii \omega) + 1} \,.
\end{align}
Now, if one can ignore the exponential suppression
factor $\exp(-2 \alpha \omega z)$ over the
entire characteristic $\omega$ integration region,
then one can approximate the interaction energy
by the very simple expression
\begin{align}
\label{nonretard}
\mathcal{E}(z) \approx & \;
-\frac{1}{4 \pi z^3} \int\limits_0^\infty
\alpha(\ii \omega) \,
\frac{\epsilon(\ii \omega) -1}%
{\epsilon(\ii \omega) + 1} =
-\frac{C_3}{z^3} \,,
\end{align}
where $C_3$ is defined in the obvious way.
This is precisely the short-range asymptotic limit
[in the expansion in powers of $z$ and $\ln(z)$)]
of the atom-surface interaction energy.
The term given in Eq.~\eqref{nonretard}
corresponds to the expression $-C_{3}/z^3$;
the leading short-range  $C_{3}$ coefficient is
otherwise listed in Eq.~(35) of Ref.~\cite{JeMo2023blog}
(it is called $C_{30}$ in Ref.~\cite{JeMo2023blog})
and in Eq.~(16.24) of Ref.~\cite{BoKlMoMo2009}.
However, if one cannot ignore the
exponential suppression (retardation) factor
$\exp(- 2 \, \alpha\,  \omega \, z)$ over the relevant
characteristic $\omega$ integration region,
then the short-range expansion breaks down,
and the atom-surface interaction energy is no
longer well approximated by Eq.~\eqref{nonretard}.
We can thus conclude that the
nonretardation approximation is valid in the
distance range
\begin{equation}
\label{cond}
z \lesssim \frac{1}{\alpha \, \omega_{\rm ch}}
\end{equation}
where $\omega_{\rm ch}$ is the {\em largest} characteristic
frequency in the problem, i.e., either in
the polarizability or in the dielectric
function of the material.

The {\em largest} characteristic
excitation frequency will typically
be obtained from the atom, not from the
solid.  Typical characteristic excitation energies
for solids are in the range of a few eV,
as is evident from the extensive tabulation
of dielectric functions in Ref.~\cite{Pa1985}.
For conductors whose dielectric function is
described by the plasma model given
in Eq.~\eqref{plasma_model},
the characteristic absorption frequency is
zero. This is evident if one
writes the expression for the plasma model
dielectric function as
$1 + \omega_{\rm pl}^2/(\omega^2 + \omega_0^2)$
with $\omega_0 = 0$.

\begin{figure}[t]
\begin{center}
\includegraphics[width=0.93\linewidth]{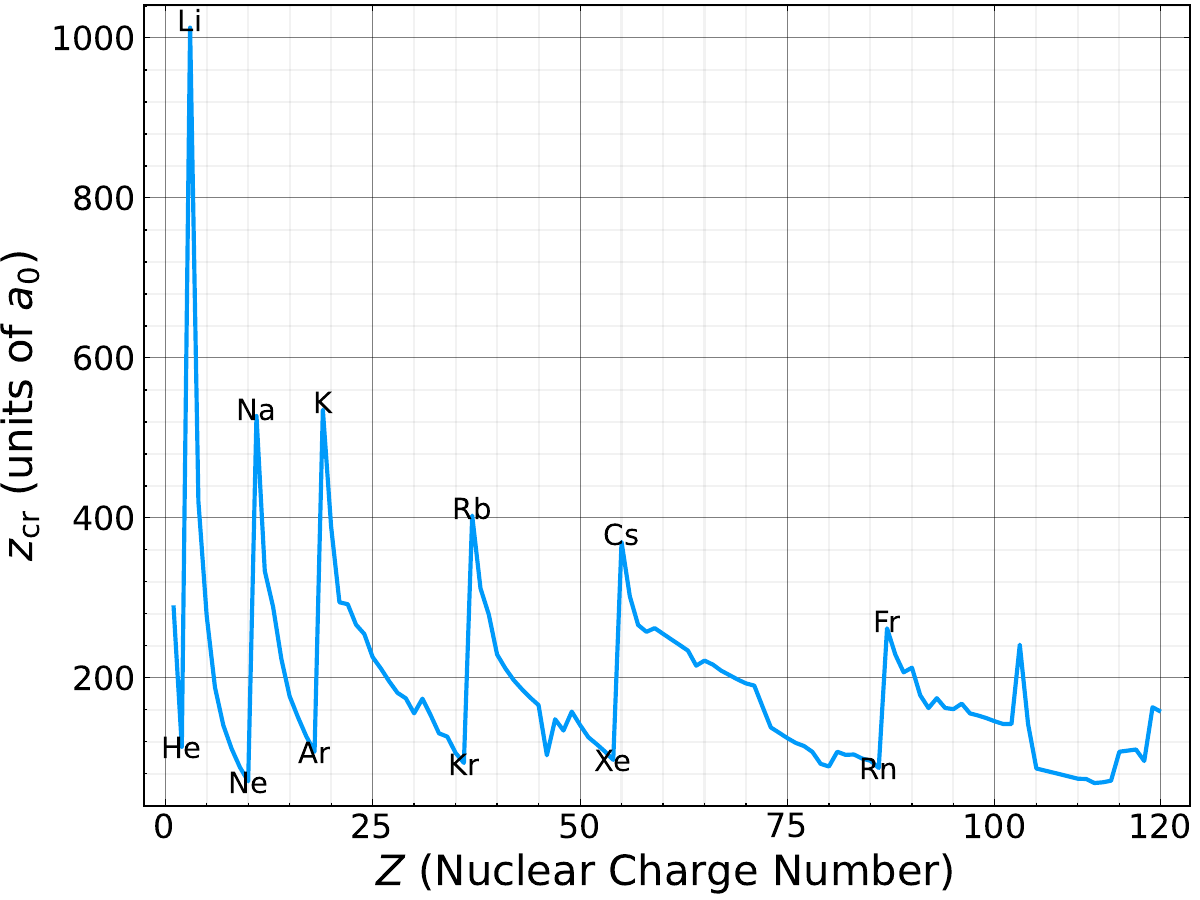}
\end{center}
\caption{\label{fig6}
Critical distance $z_{\crit}$  versus nuclear charge number
$Z$, for all elements with $1 \leq Z \leq 120$
given in Ref.~\cite{ScNa2019}.}
\end{figure}

Let us use the single-oscillator model
and define the {\em critical} distance
$z_{\crit}$ for the onset of retardation
effects to be the scale where the condition~\eqref{cond}
breaks down. This means that the critical angular frequency and its
corresponding distance scale (in atomic units) is
\begin{equation}
\label{zcritres}
\omega_{\crit} = \sqrt{ \frac{Z}{\alpha(0)} } \, {\rm a.u.} \,,
\qquad
z_{\crit} = 137 \, \sqrt{\frac{\alpha(0)}{Z}} \, {\rm a.u.}.
\end{equation}
The estimates from Eq.~\eqref{zcritres} read as
follows (using data from Ref.~\cite{ScNa2019}),
\begin{subequations}
\label{zcrit}
\begin{align}
z_{\crit}({\rm H}) =& \; 137 \,\sqrt{\frac{4.5}{1}} \, {\rm a.u.}
= 290 \, {\rm a.u.} \,,
\\
z_{\crit}({\rm He}) =& \; 137 \sqrt{\frac{1.383}{2}} \, {\rm a.u.}
= 113 \, {\rm a.u.} \,,
\\
z_{\crit}({\rm He}^*) =& \; 137 \sqrt{\frac{316}{2}} \, {\rm a.u.}
= 1720 \, {\rm a.u.} \,.
\end{align}
\end{subequations}
The correspondence (in terms of the
order-of-magnitude)
$z_{\crit}({\rm H}) \sim z_{\brek}({\rm H})$,
$z_{\crit}({\rm He}) \sim z_{\brek}({\rm He})$,
and $z_{\crit}({\rm He}^*) \sim z_{\brek}({\rm He}^*)$,
is obvious. In the latter case,
the order-of-magnitude approximation $z_{\crit}({\rm He}^*)$
is larger than $z_{\brek}({\rm He}^*)$ by about 27\,\%,
which is perfectly acceptable given the cursory nature
of the approximation.
A table of values of $z_{\crit}$
for all elements with $1 \leq Z \leq 120$
is presented in Fig.~\ref{fig6}.
Peak values are observed for alkali metals,
which typically display a very large
static polarizability.

For the case studied here,
we have $z_{\crit} > z_{\brek}$,
i.e., the onset of retardation happens a bit earlier
than predicted by the approximation $z \approx z_{\crit}$.
At $z = z_{\crit}$, we have
\begin{equation}
\exp(-2 \alpha \omega_{\crit} \, z_{\crit}) =
[ \exp(1) ]^{-2} = 0.135 \,,
\end{equation}
indicating that, for typical excitation frequencies
$\omega \sim \omega_{\crit}$ the
exponential suppression of the
integrand in Eq.~\eqref{expo_suppress}
is already very substantial at $z = z_{\crit}$.
This fact supports the observation that $z_{\brek} < z_{\crit}$.
Still, the approximation $z_{\brek} \approx
z_{\crit}$ remains a good, albeit somewhat cursory,
estimate for the transition to the retarded regime.

%
%
\section{Conclusions}
\label{sec4}

The world of atomic physics is full of surprises
in terms of nonparametric prefactors.
An example is the variation of the static
polarizability with the atomic species
(element number).
Parametrically, the static polarizabilities
of all atoms are of order $e^2 a_0^2/E_h$,
where $e$ is the electron charge,
$a_0$ is the Bohr radius, and $E_h$ is the
Hartree energy. However,
large nonparametric prefactors multiply this
estimate, two extreme cases being
metastable triplet $2{}^3 S_1$ helium with a static polarizability
of 315\,a.u., and lithium with
a static polarizability of 164\,a.u..

Another surprise, not treated here
in further detail, is the large static 
polarizability of metastable singlet $2{}^1 S_0$ helium,
which amounts to about $800$~atomic units.
However, atomic polarizabilities are not 
the only quantities leading to surprises.
As an example in a different context,
the so-called relativistic Bethe logarithm
for $S$ states of hydrogen, parametrically,
was estimated to be of order
$\alpha (Z\alpha)^4 E_h$.
After considerable efforts
by a number of groups~\cite{ErYe1965a,ErYe1965b,Er1971,Sa1981,Pa1993},
a surprising nonparametric prefactor
$\approx -31$ (for nuclear charge number $Z=1$) was confirmed to multiply the
parametric estimate, shifting theoretical
predictions for the Lamb shift
of the ground state of hydrogenlike bound 
systems~\cite{ErYe1965a,ErYe1965b,Er1971,Sa1981,Pa1993,JeMoSo1999}.
Furthermore, the nonlogarithmic prefactor 
was seen to remain numerically large
(in between $-28$ and $-31$) for nuclear
charge numbers $1 \leq Z \leq 5$.
In order to appreciate the large magnitude of the 
nonlogarithmic correction to the Lamb shift,
one observes that the 
relativistic Bethe logarithm (nonlogarithmic)
correction of order for the ground state
of hydrogenlike boron (nuclear charge number $Z=5$)
almost doubles the effect of the leading 
double logarithm 
$-\alpha (Z\alpha)^4 E_h \, \ln^2[(Z\alpha)^{-2}]$,
as is evident from 
see Eq.~(3) and Table~I of Ref.~\cite{JeMoSo1999}.

Within the context of atom-wall interactions,
we can expect the nonretarded regime
to extend furthest for those atoms
with the highest static polarizabilities
at the lowest nuclear charge numbers.
Indeed, we have demonstrated that the onset of
retardation in atom-wall interactions
depends quite significantly on the atomic species,
even if, parametrically, the estimate~\eqref{hegel}
remains valid for all.
For simple atomic systems such as hydrogen
and ground-state helium, retardation effects
set in already at distances of less than
$10\,{\rm nm} \approx 200 \, {\rm a.u.}$ in atom-wall
interactions. This result is consistent
with remarks in the text following
Eq.~(2.16) of Ref.~\cite{Ho1980}.
The breakdown of the short-range
$1/z^3$ approximation happens at distance
scales indicated in Eqs.~\eqref{zbreakSi},~\eqref{zbreakAu}
and~\eqref{zcrit}.
An explicit estimate, $z_{\crit} = 137 \, (\alpha(0)/Z)^{1/2} \,
{\rm a.u.}$, was given in Eq.~\eqref{zcritres}.

With the exception of
lithium (and metastable helium), the critical distance for
the onset of retardation effects
does not exceed $600 \, {\rm a.u.}$,
as shown in Fig.~\ref{fig6}.
An exceptional example is provided by metastable helium
where $z_{\crit}$ assumes the
exceptionally large value of $1720 \, {\rm a.u.}$,
in view of an exceptionally large
static polarizability of $315.63$ atomic units.
However, the actual breakdown distance
for metastable helium is smaller,
namely,
$1033 \, \rm{a.u.}$ and $1414 \, \rm{a.u.}$
for interactions with silicon
and gold, respectively (the latter being described
by a simple plasma model).

We can thus confirm that comparatively large $z_{\brek}$
can be expected for metastable helium,
especially for interactions with very good
conductors, an extreme example being provided
by the plasma model~\eqref{plasma_model}
for gold. However, even in this extreme case,
the nonretarded regime is limited
to about 1400\,a.u..
We conclude that the short-range approximation,
for atom-wall interactions,
breaks down much earlier
than for solid-solid
interactions~\cite{BoKlMoMo2009},
and we provide estimate for all elements
in the periodic table (see Fig.~\ref{fig6}).

%
%
\section*{Acknowledgments}

Helpful conversations with M.~DeKieviet and C.~Moore are
gratefully acknowledged.
T.~D.~and U.~D.~J.~were supported by NSF grant PHY--2110294.
C.~A.~U.~acknowledges support from NSF grant DMR--2149082.

\appendix

\section{Definition of Distance Ranges}
\label{appa}

The designations of ``short-range'' and ``long-range''
asymptotics crucially depend on the point of view. Because
the designations are not always consistent,
we here present a mini-review of this issue.

Zaremba and Kohn \cite{ZaKo1976} define ``close range'' to be the range of a
few atomic radii, commensurate with their aim to study the transition from
physisorption to the van der Waals regime; the latter is understood as the
``long-range regime'' in Ref.~\cite{ZaKo1976}.

On the other hand, Antezza {\em et al.} \cite{AnPiSt2004} define the
``long-range regime'' as the limit of very large separations far beyond the
validity of van der Waals and Casimir-Polder interactions.  This limit is
characterized by a very-long range nonretarded tail proportional to $1/z^3$,
which is due to effects described by thermal field theory (contributions of the
first Matsubara frequency) and vanishes at zero temperature.  The numerical
coefficient of this  extreme $1/z^3$ long-range tail is very small (see
Eq.~(17) of Ref.~\cite{AnPiSt2004}) and we do not consider it here.

Thus, from the viewpoint of Ref.~\cite{AnPiSt2004}, the extreme short
range is the regime of less than ten atomic radii, where the discretization of
the crystal surface starts to play a role. The short-range regime is the
$1/z^3$ nonretarded (van der Waals) range. The $1/z^4$ Casimir-Polder
interactions then define the long range regime.
This is the viewpoint we also adopt in the present paper.

For completeness, let us also say a few words about the limit of very close
approach to the surface. Zaremba and Kohn \cite{ZaKo1976} showed that in this
limit the van der Waals interaction becomes $-C_3/(z-z_0)^3$, where  $z_0$ is a
parameter, of order unity in atomic units, which can be calculated separately
or obtained experimentally.  For dielectric solids, the position of the
reference plane is well approximated by $z_0 \approx d/2$, where $d$ is the
distance between layers of the substrate \cite{Ho1980,TaRa2014}.

In other investigations~\cite{SiAmGrAn2012,TaRa2014},
van-der-Waals corrected density-functional theory (DFT) is used
in order to calculate the adsorption energies
of atoms on surfaces (e.g. rare gases on noble metals).
The quadrupole correction is routinely taken
into account in this procedure (see Table~III of
Ref.~\cite{TaRa2014} and Ref.~\cite{Je2024multipole}),
and the van der Waals energy is added to the
contact energy at the equilibrium position of the atom
in the immediate vicinity of the surface,
the latter being calculated with the use of DFT
(see Table~III of Ref.~\cite{TaRa2014},
and also Ref.~\cite{GrAnEhKr2010} for a
general discussion of van-der-Waals corrected DFT).
This procedure is consistent with remarks made
after Eq.~(2.39) of Ref.~\cite{ZaKo1976}, where the authors
stress that their approach should be valid for the
region of physisorption (i.e., for the
range in between 4 and 7 atomic units).

%
%
\section{Parameters of the Helium Calculation}
\label{appb}

The $a$, $b$ and $c$ parameters of the exponential 
basis functions $\exp(-a r_1 - b r_2 - c r_{12})$ set are
chosen in the same way for the reference ground state
and the triple metastable excited reference state.
One notes that our choice is different from,
say, the basis set indicated in Eq.~(18) of Ref.~\cite{AzBeKo2018}.
Through experimentation, we found 
it numerically favorable to implement a linear dependence
of the exponents of the basis functions in the
exponential basis with the index of the function.

Specifically, we use a Cartesian approach~\cite{YePaPa2021}
and write the following expressions
for the singlet ${}^1 S$, triplet ${}^3 S$, singlet ${}^1 P$,
and triplet ${}^3 P$ states,
\begin{subequations}
\begin{align}
\psi_{n {}^1 S}(\vec r_1, \vec r_2) = & \;
\sum_{j=1}^{j_{\rm max}} c_{n{}^1S,j} \, 
f_{n{}^1S,j}(r_1, r_2, r_{12}) \,,
\\
f_{n{}^1S,j}(r_1, r_2, r_{12}) =& \;
\exp(-a_j r_1 - b_j r_2 - c_j r_{12})
\\[0.1133ex]
& \; + \exp(-b_j r_1 - a_j r_2 - c_j r_{12})  \,,
\nonumber\\
\psi_{n {}^3 S}(\vec r_1, \vec r_2) = & \;
\sum_{j=1}^{j_{\rm max}} c_{n{}^3S,j} \,
f_{n{}^3S,j}(r_1, r_2, r_{12}) \,,
\\
f_{n{}^3S,j}(r_1, r_2, r_{12}) =& \;
\exp(-a_j r_1 - b_j r_2 - c_j r_{12})
\\[0.1133ex]
& \; - \exp(-b_j r_1 - a_j r_2 - c_j r_{12}) \,,
\nonumber\\
\psi^i_{n {}^1 P}(\vec r_1, \vec r_2) = & \;
\sum_{j=1}^{j_{\rm max}} c_{n {}^1P,j} \, 
f^i_{n{}^1P,j}(r_1, r_2, r_{12}) \,,
\\
f^i_{n{}^1P,j}(r_1, r_2, r_{12}) =& \;
x_1^i \exp(-a_j r_1 - b_j r_2 - c_j r_{12})
\\[0.1133ex]
& \; + x_2^i \exp(-b_j r_1 - a_j r_2 - c_j r_{12}) \,,
\nonumber\\
\psi^i_{n {}^3 P}(\vec r_1, \vec r_2) =& \;
\sum_{j=1}^{j_{\rm max}} c_{n {}^3P,j} \,
f^i_{n{}^3S,j}(r_1, r_2, r_{12}) \,,
\\[0.1133ex]
f^i_{n{}^3P,j}(r_1, r_2, r_{12}) =& \;
x_1^i \exp(-a_j r_1 - b_j r_2 - c_j r_{12})
\\[0.1133ex]
& \; 
- x_2^i \exp(-b_j r_1 - a_j r_2 - c_j r_{12}) \,.
\nonumber
\end{align}
\end{subequations}
The $c_{n {}^{2S+1} L, j}$ coefficients depend on the 
principal quantum number $n$, the total spin $S$,
and the total orbital angular momentum $L$.
The basis states $f_{n {}^{2S+1} L, j}(r_1, r_2, r_{12})$
multiply the coefficients.
Within the manifold of states of specified symmetry,
one calculates the Hamiltonian matrix
within the basis spanned by the $f_{n {}^{2S+1} L, j}(r_1, r_2, r_{12})$,
and the overlap matrix of the same basis states,
using the approach outlined in Ref.~\cite{AzBeKo2018} and Chap.~13 of 
Ref.~\cite{JeAd2022book}.
Let now $\mathbbm{S}_{j j'} = \langle j | j' \rangle$
be the overlap matrix and $\mathbbm{H}_{j j'} = \langle j | H_{\rm NR} | j' \rangle$,
where $H_{\rm NR}$ is the Schr\"{o}dinger Hamiltonian 
of helium in the nonrelativistic approximation (see Chap.~13 of Ref.~\cite{JeAd2022book}).
The diagonalization of the effective Hamiltonian 
$\mathbbm{H}_{\rm eff} = \mathbbm{S}^{-1/2} \, \mathbbm{H} \,  \mathbbm{S}^{-1/2}$
leads to discrete states; the first of which describe low-lying 
bound states, while the higher excited states 
with positive energy eigenvalues above the ionization threshold 
serve to describe the continuum spectrum, in terms 
of a pseudospectrum of discrete states.
The summation over $j$ goes from $j=1$ to $j_{\rm max}$.
where a value of $j_{\rm max} = 512$ turns out to be fully 
sufficient for our purposes here.
Let us relate the summation index $j = (n_{\rm max})^3$ 
to three summation indices,
$k$, $\ell$, $m$, all of which go from $1$ to $n_{\rm max} = 8$
(in our calculation),
\begin{subequations}
\label{mapping}
\begin{align}
k \in & \; \{1, \cdots, n_{\rm max} \} \,,
\qquad
\ell \in \{1, \cdots, n_{\rm max} \} \,,
\\[0.1133ex]
m \in & \; \{1, \cdots, n_{\rm max} \} \,,
\qquad
j_{\rm max} = (n_{\rm max})^3 \,,
\\[0.1133ex]
j = & \; k-1 + n_{\rm max} (\ell - 1) + (n_{\rm max})^2 (m - 1) + 1
\end{align}
\end{subequations}
This generates a basis of $(n_{\rm max})^3$ functions.
The relations~\eqref{mapping} 
define a unique mapping $(k, \ell, m) \leftrightarrow j$.
Given $j$, one can calculate 
$k = k(j)$, $\ell = \ell(j)$, and $m = m(j)$,
using $m = ( j - {\rm mod}[j-1, (n_{\rm max})^2] - 1 )/(n_{\rm max})^2 + 1$,
$\ell = ( j - (m-1) (n_{\rm max})^2 - 
{\rm mod}[j - (m - 1) (n_{\rm max})^2 - 1, n_{\rm max}] - 1 )/n_{\rm max} + 1$,
and $k = j - (m - 1) (n_{\rm max})^2 - (\ell - 1) \, n_{\rm max}$.
The parameters are chosen to depend linearly on $k$, $\ell$, and $m$,
in such a way as to avoid degeneracies in the basis
(hence using prime numbers),
\begin{subequations}
\begin{align}
a_j =& \; a_{k(j)} = \frac{\sqrt{3}}{10} \, (k+1) \,,
\\[0.1133ex]
b_\ell =& \; b_{\ell(j)} = \frac{9 \, (\sqrt{17} - \sqrt{5})}{100} \,
(\ell+1) \,,
\\[0.1133ex]
c_m =& \; c_{m(j)} = \frac{1}{10} \, (m+1) \,.
\end{align}
\end{subequations}
This choice involves basis functions
of the type $\exp(-a r_1 - b r_2 - c r_{12})$
with numerically large coefficients
$a = a_k$, $b = b_\ell$, and $c = c_k$, 
achieved for $k, \ell, m = n_{\rm max}$.
Consequently, steep exponential decay is realized for
some of the basis functions, which is helpful 
in a suitable description of the region near
the cusp, $r_{12} = 0$.
For low values of $k, \ell, m$, we achieve a good 
sampling of the large-distance region,
which is important for a good description of oscillator
strengths.

With very modest computational effort,
this approach reproduces other data~\cite{DrYa1992,YaBa1998,Dr1999,Dr2005,AzBeKo2018}
for low-lying energy levels
of helium, and for the static and dynamic polarizability of
helium (within the nonrelativistic approximation),
to better than one permille~\cite{PaSa2000}.
Specifically, we obtain 
a value of $1.383192 \, {\rm a.u.}$ for the ground-state static polarizability,
which compares well with~\cite{PaSa2000,Dr2005},
and an oscillator strength of $0.276167 \, {\rm a.u.}$
for the $1 {}^1 S$--$2 {}^1 P$ oscillator strength,
which compares well with Ref.~\cite{Dr2005}.
For the static polarizability of the metastable triplet 
state, we obtain a value of 
$315.63 \, {\rm a.u.}$, which compares well with Ref.~\cite{ScNa2019}.
For the $2 {}^3 S$--$2 {}^3P$ oscillator strength,
we obtain a value of $0.5391  \, {\rm a.u.}$,
which compares well with Ref.~\cite{We1967}.

%
%
\section{Intrinsic Silicon}
\label{appc}

In this appendix, we provide
a brief review of the Clausius--Mossotti
fits recently employed
in Ref.~\cite{MoEtAl2022} for intrinsic silicon.
We also take the opportunity to correct a few
typographical errors.
From Ref.~\cite{MoEtAl2022},
we recall the
Lorentz–Dirac master function, as follows:
\begin{equation}
\label{master}
f(T_\Delta, \omega) = \sum_{k=1}^{k_{\rm max}}
\frac{a_k(\omega_k^2 - \ii \gamma_k'\omega)}%
{\omega_k^2 - \omega^2 - i\omega \gamma_k} \,,
\end{equation}
where $T_\Delta = (T-T_0)/T_0$ and $T_0$ is the room temperature.
In Refs.~\cite{OuCa2003} and~\cite{LaDKJe2010pra},
inspired by the Clausius–Mossotti relation, the dielectric ratio
\begin{equation}
\label{defrho}
\rho(T_\Delta,\omega) =
\frac{\epsilon(T_\Delta,\omega)-1}{\epsilon(T_\Delta,\omega)+2}
\doteq f(T_\Delta, \omega)
\end{equation}
was fitted to a functional form corresponding to the
master function. That is to say, one fits
\begin{equation}
\label{CMfit}
\epsilon_{\rm CM}(T_\Delta,\omega)  \doteq
\frac{1 + 2f(T_\Delta,\omega)}{1-f(T_\Delta,\omega)} \,.
\end{equation}
We now take the opportunity to correct
two unfortunate typographical errors
in Ref.~\cite{MoEtAl2022},
Equation~(8) of Ref.~\cite{MoEtAl2022} misses an
opening curly parenthesis in the numerator,
\begin{equation}
\label{cCMFit}
\rho(T_\Delta,\omega)
= \sum_{k=1}^{k_{\rm max}}
\frac{a^\CM_k(T_\Delta) \, \{ [ \, \omega^\CM_k(T_\Delta) \, ]^2 -
\ii \, \gamma^{\prime\CM}_k(T_\Delta) \, \omega \} }
{[ \, \omega^\CM_k(T_\Delta) \, ]^2
- \omega^2
- \ii \, \omega \, \gamma^\CM_k(T_\Delta)} ,
\end{equation}
while Eq.~(9) of Ref.~\cite{MoEtAl2022} has
a typographical in the last term of the numerator;
one needs to replace
$\omega^4 \to \left[\omega^\CM_k(T_\Delta)\right]^4$,
\begin{multline}
\label{eq:reFitCM}
\mathrm{Re}[\rho_\CM(T_\Delta, \omega)] =
\sum_{k=1}^{k_{\rm max}} a^\CM_k(T_\Delta)
\\
\times
\frac{ \omega^2 \left[ \, \gamma^\CM_k(T_\Delta) \,
\gamma_k^{\prime\CM}(T_\Delta) -
[\omega^\CM_k(T_\Delta)]^2 \, \right] +
[\omega^\CM_k(T_\Delta)]^4 }%
{(\omega^2-[\omega^\CM_k(T_\Delta)]^2)^2+ \omega^2\,
\left[\gamma^\CM_k(T_\Delta)\right]^2}\,,
\end{multline}
while the imaginary part is
\begin{multline}
\label{eq:imFitCM}
\mathrm{Im}[\rho_\CM(T_\Delta, \omega)] =
\sum_{k=1}^{k_{\rm max}} a^\CM_k(T_\Delta) \, \omega \,
\\
\times \frac{ \omega^2 \gamma_k^{\prime\CM}(T_\Delta)
+ \left\{ \gamma^\CM_k(T_\Delta) -
\gamma_k^{\prime\CM}(T_\Delta) \right\} [\omega^\CM_k(T_\Delta)]^2 }%
{ \{ \omega^2-[\omega^\CM_k(T_\Delta)]^2 \}^2 +
\omega^2\, \left[ \, \gamma^\CM_k(T_\Delta) \, \right]^2}.
\end{multline}

\begin{table}
\caption{\label{table1}  Parameters for the CM fit are
indicated for the real and for the
imaginary parts of the dielectric function for silicon,
as given in Eqs. (\ref{eq:reFitCM})
and (\ref{eq:imFitCM}), at room temperature
($T_\Delta=0$ in the notation of Ref.~\cite{MoEtAl2022}).
Here, $a_{1,2}$ are
dimensionless and $\omega_{1,2}$, $\gamma_{1,2}$ and $\gamma'_{1,2}$ are in
units of $E_h/\hbar$. The values are adapted from Tables I and II of
Ref.~\cite{MoEtAl2022},
with minor adjustments of the entries marked by $^*$ (see text).  }
\begin{ruledtabular}
\begin{tabular}{ccccc}
$k$ & $a_k$ & $\omega_k$ & $\gamma_k$ & $\gamma'_k$\\
\hline
1 & 0.004943 & 0.1293 & 0.01841 & 0.1306 \\
2 & 0.7709 & 0.3117 & 0.101$^*$ & 0.0968$^*$
\end{tabular}
\end{ruledtabular}
\end{table}

We note that the imaginary part of the dielectric function should be strictly
positive for real, positive frequencies, due to causality (see Chap.~6 of
Ref.~\cite{Je2017book}).  The region near $\omega \approx 0$ of the CM
fit for the imaginary part of $\mbox{Im}[\epsilon(\omega)]$ has a positive
second derivative. This means that, between the frequency points $\omega = 0$
and $\omega = \omega_{\rm min}$, where $\omega_{\rm min}$ is the smallest
frequency for which the dielectric function has been measured, one has a
``gap'' where any fitting function runs the risk of ``undershooting'' the line
${\rm Im}[\epsilon(\omega)] = 0$, and the positive second derivative
compensates a negative first derivative at $\omega = 0$ to match the points of
smallest frequency of the fit.

\begin{figure}[t]
\begin{center}
\includegraphics[width=0.99\linewidth]{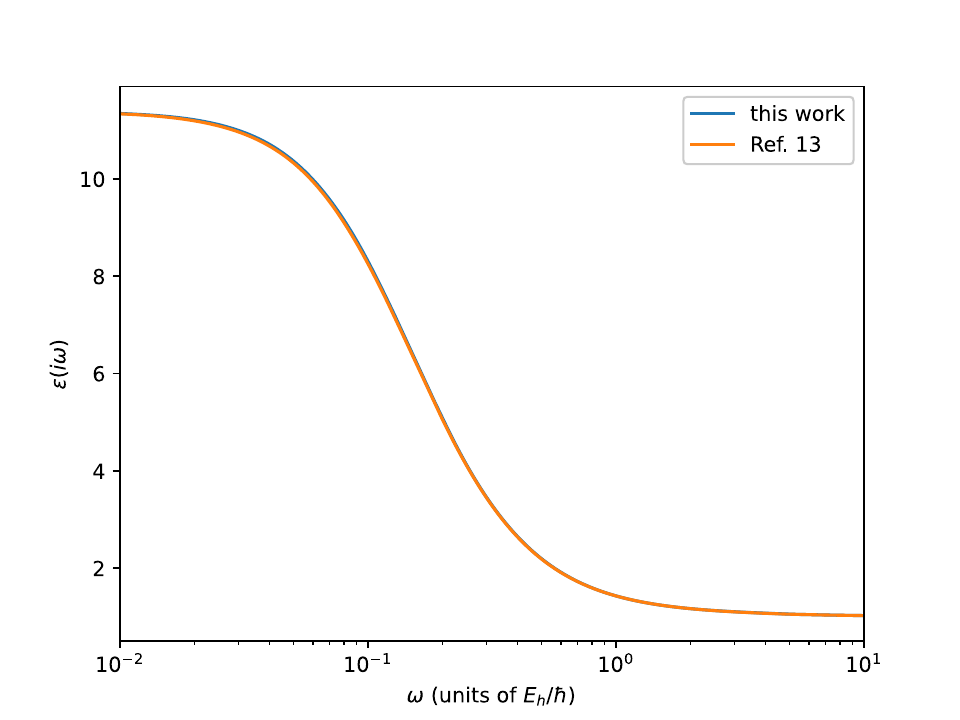}
\end{center}
\caption{\label{fig7}
Dielectric function of silicon at room temperature for imaginary frequency
argument, $\epsilon(\ii \omega)$. The two curves compare the CM fit using the
parameters of Ref.~\cite{MoEtAl2022}  and the adjusted parameters of Table
\ref{table1}. The abscissa indicates the 
angular frequency in atomic units, i.e,
the numerical value of $\hbar \omega/E_h$, where 
$E_h$ is the Hartree energy.}
\end{figure}

Indeed, the fitting parameters that were given in Ref. \cite{MoEtAl2022} led to
spurious negative imaginary parts under some circumstances, but these were so
small as to be statistically insignificant (less than 0.2 \% of the total
$\epsilon(\omega)$).  Table \ref{table1} gives the best CM fitting parameters
at room temperature, where some entries were slightly adjusted compared to the
values given in Ref.~\cite{MoEtAl2022} (but still within the error bars of the
fit) to ensure overall positivity of $\mbox{Im}[\epsilon(\omega)]$.
Figure~\ref{fig7} shows  $\epsilon(i\omega)$, comparing the adjusted CM fitting
parameters of Table \ref{table1} with the original parameters of
Ref.~\cite{MoEtAl2022}. The two curves are essentially on top of each other, with a
maximal difference of 0.6 \%.

%
%
\section{TRK Sum Rule for Metastable States}
\label{appd}

The TRK sum rule~\cite{ReTh1925zahl,Ku1925}
is instrumental in deriving
the correct asymptotic form of the dynamic polarizability
at large imaginary frequency. It states that the
sum over all oscillator strengths is
equal to the number of electrons $Z$ of the atom.
According to Eq.~(61.1) of Ref.~\cite{BeSa1957},
it is valid for an arbitrary
(e.g., metastable) reference state $| \psi_m \rangle$,
\begin{equation}
\sum_n f_{nm} = Z \,,
\end{equation}
where $n$ sums over all quantum numbers of the
system (not just the principal ones).
This is confirmed in Eq.~(22) of Ref.~\cite{DrDa1970}
for excited singlet and triplet 
reference states which are embedded in a continuum.
When comparing to Ref.~\cite{DrDa1970},
one notes that the authors of Ref.~\cite{DrDa1970}
use, in some parts of their investigations, a somewhat nonstandard
redefinition of the oscillator strengths, 
adapted to different magnetic projections.
Here, we adopt the standard definition of the 
oscillator strength, which entails an 
average over the magnetic projections 
of the reference state $| m \rangle$,
and a summation over the magnetic projections
of the virtual state $| n \rangle$
(see Sec.~5.5.3 of Ref.~\cite{JeAd2022book} and
Ref.~\cite{Hi1982}).
In view of the relation
\begin{equation}
\alpha(\ii \omega) = \sum_n
\frac{f_{nm}}{\omega_{nm}^2 + \omega^2} \,,
\end{equation}
the TRK sum rule determines the
asymptotic behavior of the polarizability
for large $\omega$ (the energy difference
in atomic units
of the virtual and the reference state is $\omega_{nm} =
\omega_n - \omega_m$).
Here, we present a derivation which,
in contrast to Eq.~(11.10) of Ref.~\cite{BeJa1986},
is valid for a system with an arbitrary
number $Z$ of electrons, and
for an arbitrary reference state.  The Hamiltonian is
\begin{equation}
H = \sum_a \left(\frac{\vec p_a^{\,2}}{2} - \frac{Z}{r_a}\right) +
\sum_{a<b} \frac{1}{r_{ab}} \,,
\end{equation}
where $a$ and $b$ sum over the electrons,
$r_a$ is the electron-nucleus distance, and
$r_{ab}$ is the interelectron distance.
Indices $a,b,c,d \in \{1,\ldots,Z\}$ enumerate the bound electrons.
The (dipole) oscillator strength for the
excitation to the states $| \psi_n \rangle$ is
is
\begin{equation}
f_{nm} = \frac23 \, (E_n - E_m)
\, | \langle \psi_n | \sum_c \vec r_c | \psi_m \rangle |^2 \,,
\end{equation}
where we reemphasize the 
average over the magnetic projections
of the reference state $| m \rangle$,
and a summation over the magnetic projections
of the virtual state $| n \rangle$.
The sum over oscillator strengths can be written
as follows, in operator notation,
\begin{equation}
\sum_n f_{nm} = \frac23 \,
\langle \psi_n | \left( \sum_c \vec r_c \right) \, (H - E_m)
\left( \sum_d \vec r_d \right) | \psi_m \rangle  \,,
\end{equation}
where the sum over $n$ includes the continuum.
We now make use of the well-known
operator identity
\begin{equation}
A B A = \frac12 A^2 B + \frac12 B A^2 +
\frac12 \, [A, [B, A]] \,,
\end{equation}
where $A = \sum_c \vec r_c$ and $B = H - E_m$.
Because $| \psi_m \rangle$ is an eigenstate of the
Hamiltonian, we have
$(H - E_m) | \psi_m \rangle = 0$ and thus
\begin{multline}
\sum_n f_{nm} = \frac13 \,
\langle \psi_m | \left[ \sum_c \vec r_c,
\left[ H - E_m, \sum_d \vec r_d
\right] \right] | \psi_m \rangle
\\
=
-\frac{\ii}{3} \,
\langle \psi_m | \left[ \sum_c \vec r_c,
\sum_d \vec p_d \right] | \psi_m \rangle
\\
= -\frac{\ii}{3}
\langle \psi_m | \sum_c [\vec r_c, \vec p_c] | \psi_m \rangle
= Z \,.
\end{multline}
This derivation confirms that the TRK sum rule remains
valid for metastable excited states and justifies
our parameters for the single-oscillator model
of the dynamic polarizability of metastable triplet
helium, used in Sec.~\ref{sec3}.
Recently, generalizations of the TRK sum rule suitable
for the treatment of dipole recoil terms which occur
in recoil-induced contributions to the shake-off probability
following beta decay have been discussed in Ref.~\cite{ScDr2015}.
Other generalizations with respect to multipole
polarizabilities have been considered~\cite{YaZhZh2000,ZhZhYa2006}.
Their derivation follows the ideas underlying the
above considerations.

\end{document}